\documentclass[twocolumn,twoside,slac_two]{revtex4}
\usepackage{graphicx}
\usepackage{fancyhdr}
\usepackage{hyperref}
\hypersetup{
    colorlinks=true,
    urlcolor=magenta,
}
\begin{document}

\title{Effect of pressure and temperature corrections on muon flux variability at ground level and underground}


\author{M. Savic, A. Dragic, N. Veselinovic, V. Udovicic, R. Banjanac, D. Jokovic, D. Maletic}
\affiliation{Institute of Physics, Belgrade, Pregrevica 118, Serbia}

\begin{abstract}
In Low Background Laboratory at Institute of Physics Belgrade, plastic scintillators are used to continuously monitor flux of the muon component of secondary cosmic rays. Measurements are performed on the surface as well as underground (25 m.w.e depth). Temperature effect on muon component of secondary cosmic rays is well known and several methods to correct for it are already developed and widely used. Here, we apply integral method to calculate correction coefficients and use GFS (Global Forecast System) model to obtain atmospheric temperature profiles. Atmospheric corrections reduce variance of muon flux and lead to improved sensitivity to transient cosmic ray variations. Influence of corrections on correlation with neutron monitor data is discussed.
\end{abstract}

\maketitle

\thispagestyle{fancy}

\begin{figure}[h]
\centering
\includegraphics[width=80mm]{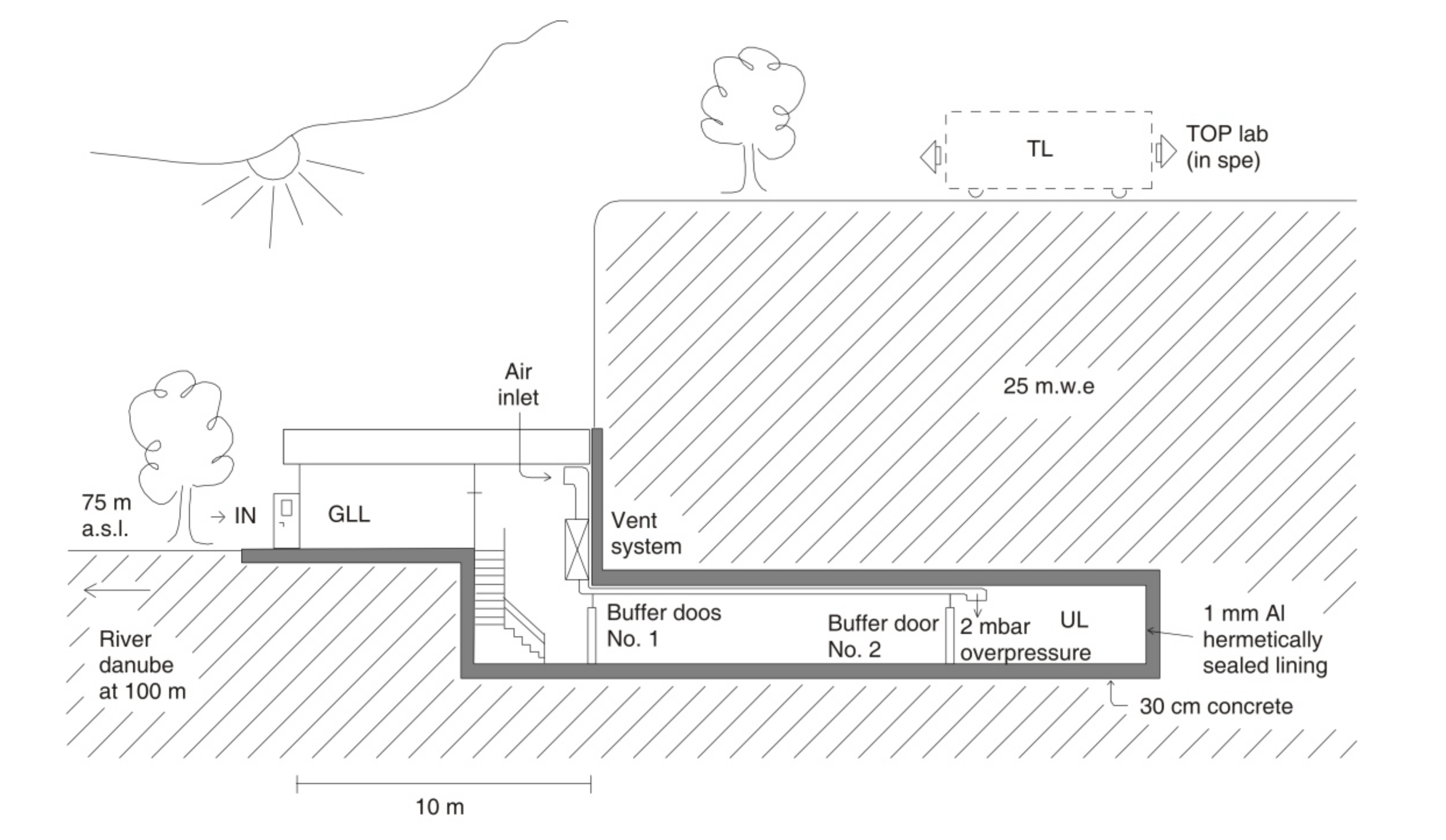}
\caption{Layout of the Low Background Laboratory.} 
\label{LBL_layout}
\end{figure}

Belgrade Low Background Laboratory (LBL) is located at Institute of Physics, Belgrade and consists of two interconnected spaces, a ground level laboratory (GLL) and a shallow underground one (UL) [Fig.~\ref{LBL_layout}]. GLL is at 75 meters above sea level while UL is dug under a 10 meter cliff and has a 12 meters of loess soil overburden (25 meters of water equivalent)~\cite{ref-1}. Geographic latitude for the site is 44.86° and longitude is 20.39° while geomagnetic rigidity cutoff is 5.3 GV. 

\begin{figure}[h]
\centering
\includegraphics[width=40mm]{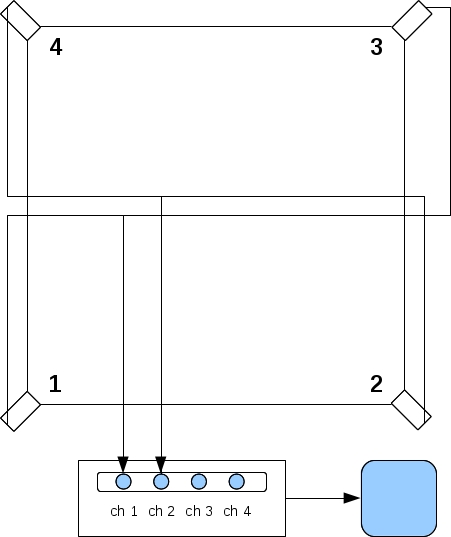}
\caption{Experimental setup scheme.} 
\label{Experimental_setup}
\end{figure}

Experimental setup consists of two identical sets of detectors and read out electronics, one situated in GLL and the other in UL. Each setup utilizes a plastic scintillator detector with dimensions 100cm x 100cm x 5cm (Amcrys-H, Kharkov, Ukraine) equipped with 4 PMTs (Hammamatsu R1306) directly coupled to the corners [Fig.~\ref{Experimental_setup}]. Flash ADC (CAEN type N1728B) with 10ns sampling are used for read out~\cite{ref-1}.

\begin{figure}[h]
\centering
\includegraphics[width=60mm]{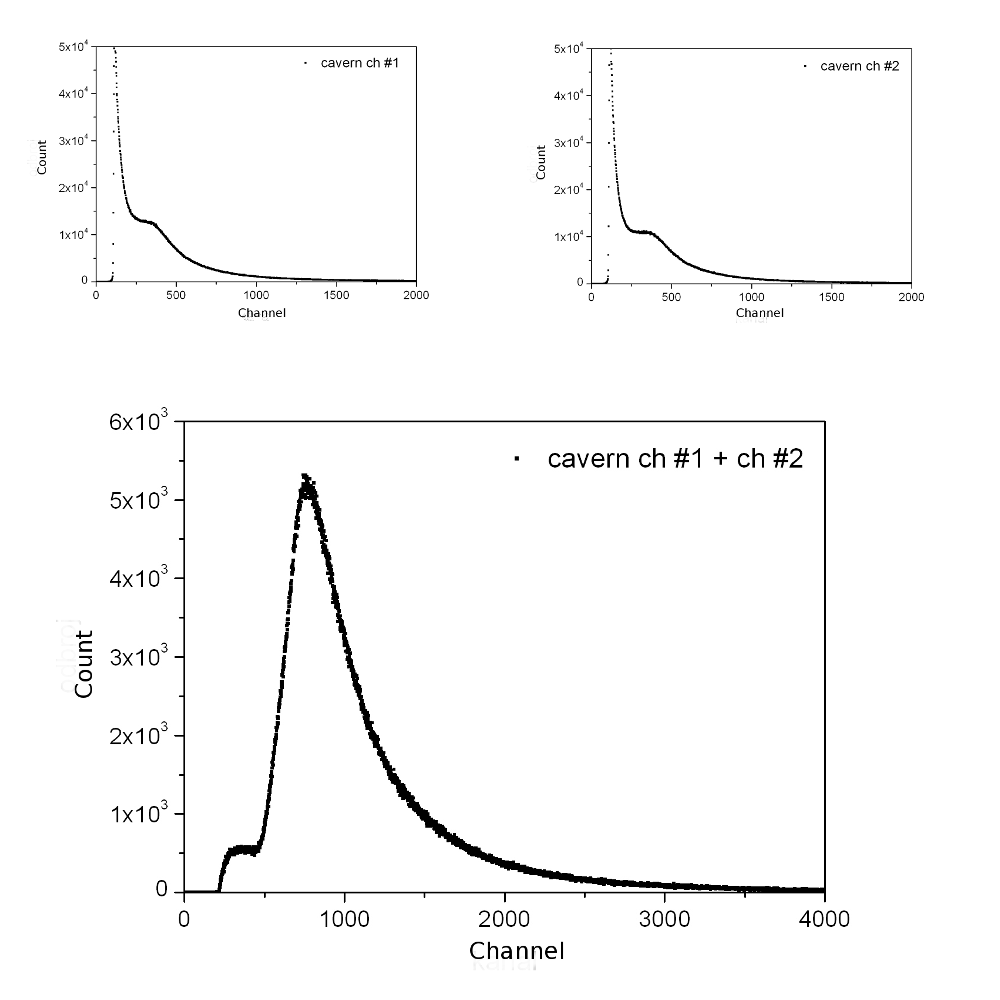}
\caption{Single summed diagonal and coincidence spectra.} 
\label{Spectra}
\end{figure}

Preamplifier outputs of two diagonally opposing PMTs are summed and fed to a single FADC input thus engaging two inputs of the FADC for two such diagonal pairings. Signals recorded by the two inputs are coincided in offline analysis, resulting in coincidence spectrum which is then used to determine the integral count [Fig.~\ref{Spectra}]. This procedure almost completely eliminates low-energy environmental background leaving only events induced by cosmic ray muons and muon related EM showers~\cite{ref-1}.

\section{Significance of meteorological effects}
Meteorological effects on muon component of secondary cosmic rays are well known, with pressure and temperature effect being most dominant~\cite{ref-2}. Correcting for these effects noticeably increases data usefulness, especially increasing sensitivity to periodic and aperiodic variations of non-atmospheric origin (variations of primary cosmic rays, different heliospheric processes, etc.)

In Belgrade Low Background Laboratory continual measurements utilizing described setup started in April of 2008 for the GLL and in November of 2008 for the UL, and with some interruptions are still ongoing. Base time resolution for integrated count is 5 minutes but time resolution of 1 hour is also often used in analysis. Link to Belgrade cosmic ray station can be found on the following address: http://www.cosmic.ipb.ac.rs/.

\subsection{Pressure effect}
Barometric effect is defined by the following equation:

\begin{equation}\label{eq:units}
\left({\delta I\over I}\right)_P = \beta \cdot \delta P
\end{equation}

where $\delta I \over I$ is the normalized variation of muon flux intensity, $\beta$ is barometric coefficient and $\delta P$ is pressure variation. Pressure variation is calculated as $\delta P = P - P_B$ , where $P$ is current pressure and $P_B$ is base pressure value~\cite{ref-4}.
 
\begin{figure}[h]
\centering
\includegraphics[width=80mm]{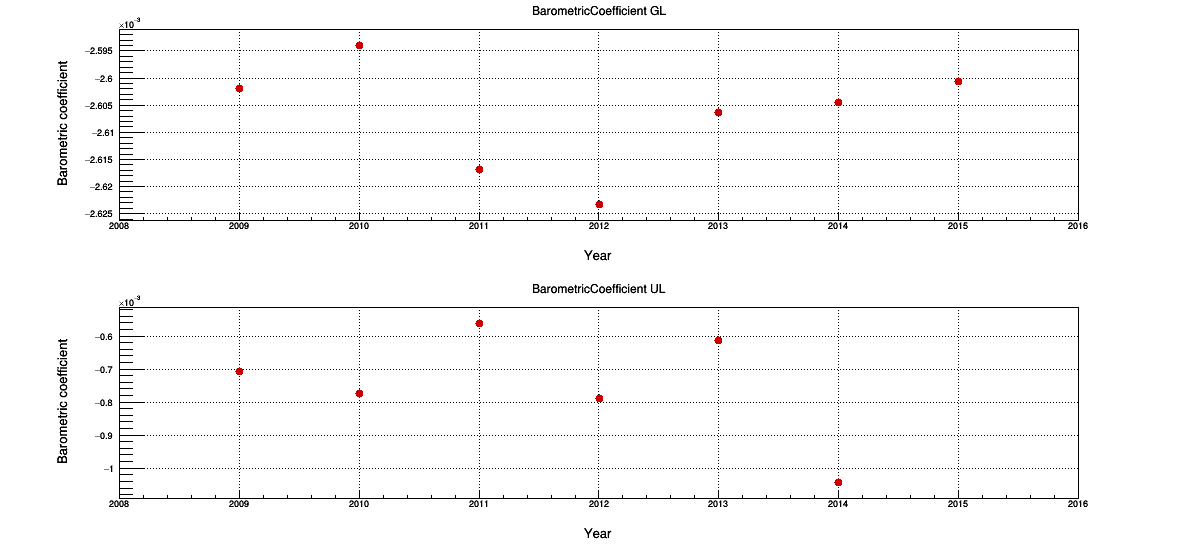}
\caption{Yearly values for barometric coefficient for GLL and UL.} 
\label{Barometric_coefficients}
\end{figure}

Since no in situ pressure measurement was performed prior to 2015, current pressure values have mostly been acquired from official meteorological measurements performed by Republic Hydrometeorological Service of Serbia as well as from Belgrade airport meteorological measurements. In all, data from 5 different stations were used. All pressure data was normalized to Belgrade main meteorological station. Stations were sorted according to geographical proximity and consistence of data. Unique pressure time series was composed by using data from the first station with available pressure entries for a given hour. Linear interpolation was then performed and pressure values were sampled with 5 minute step.
Normalized variation of muon flux intensity vs. pressure variation was plotted for each year. Only data for the 5 geomagnetically most quiet days of each month were taken into account (selected from International Quiet Days list). Barometric coefficient for each year was determined from linear fits of these plots [Fig.~\ref{Barometric_coefficients}].

\subsection{Temperature effect}
Temperature effect on hard muons is well known ~\cite{ref-2} and there are several methods developed to describe and correct for it. Method we used was integral method, where normalized variation of muon flux dependence on temperature variation is described as: 

\begin{equation}\label{eq:units}
\left({\delta I\over I}\right)_T = \int_{0}^{h_0} \alpha (h) \cdot \delta T(h) \cdot dh
\end{equation}

$\alpha (h)$ being temperature coefficient density and temperature variation calculated as
$\delta T = T - T_B$, where $T$ is current temperature and $T_B$ is base temperature value~\cite{ref-3}.

To correct for temperature effect using formula above it is necessary to have most complete information about atmospheric temperature profile for a given geographical location as well as to know temperature coefficient density function. Temperature profile measurements performed by local meteorological service are not done on consistent basis but more detailed information is available from meteorological models. One such model is GFS (Global Forecast System) that, among other data, provides temperatures for 25 isobaric levels for a given geographical location with latitude/longitude precision of 0.5 degrees ~\cite{ref-3}. 

\begin{figure}[h]
\centering
\includegraphics[width=80mm]{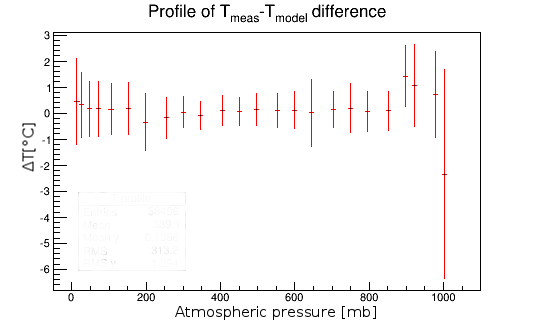}
\caption{Distribution of difference between modelled temperatures and temperatures measured by meteorological balloons above Belgrade (where such data was available).} 
\label{Profile_sigma}
\end{figure}

Measured and modelled values seem to be in fairly good agreement [Fig.~\ref{Profile_sigma}] except for the lowest isobaric level. That is why for this level temperature from local meteorological stations was used, treated in the same manner as described for local pressure data . Time resolution for modelled temperatures is 6 hours so interpolation was performed using cubic spline ~\cite{ref-3} and temperature values were sampled in 5 minute steps.

\begin{figure}[h]
\centering
\includegraphics[width=80mm]{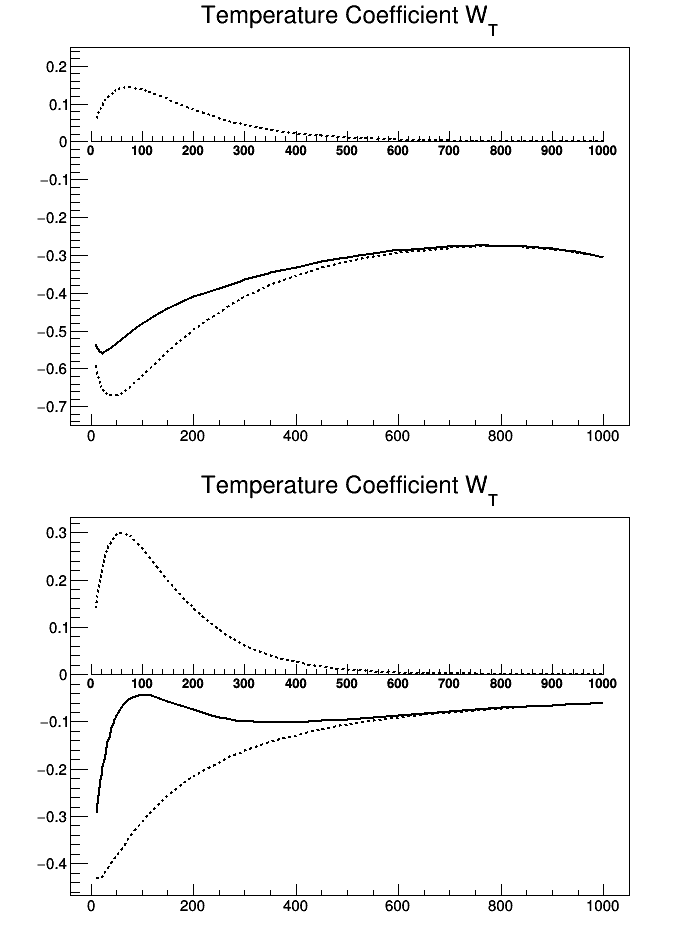}
\caption{Temperature coefficient density functions for ground level (above) and depth 25 m.w.e. (below).} 
\label{Coefficients}
\end{figure}

Temperature density functions [Fig. \ref{Coefficients}] are calculated according to procedure described in ~\cite{ref-2}. 

\section{Results}
\subsection{PT corrected time series}

\begin{figure}[h]
\centering
\includegraphics[width=70mm]{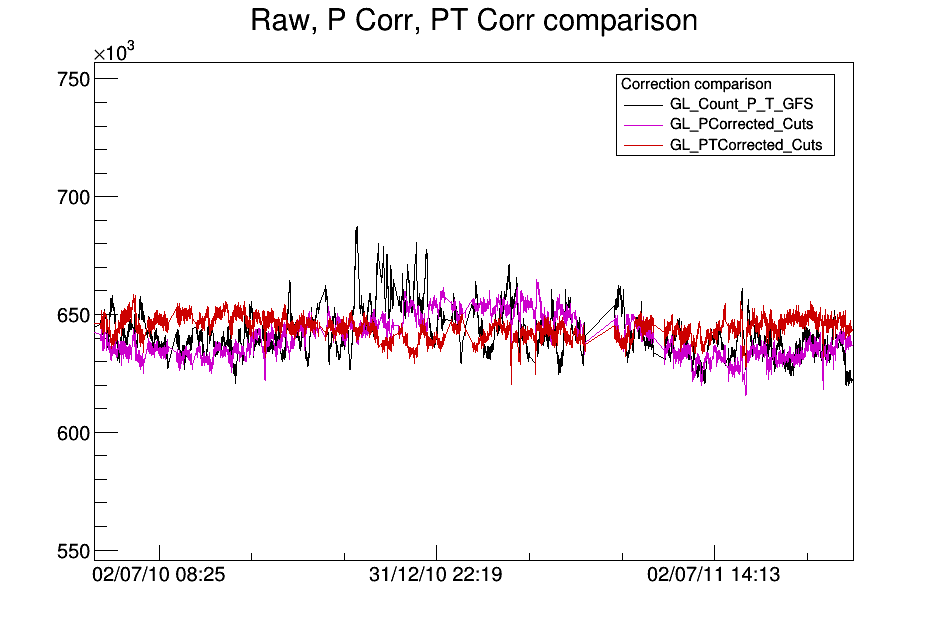}
\caption{GLL raw (black), pressure corrected (magenta) and PT corrected (red) muon count time series for a selected period.} 
\label{PT_TimeSeries}
\end{figure}

It would seem that pressure correction successfully removes aperiodic pressure induced 
fluctuations while temperature correction most significantly affects annual variation induced by atmospheric temperature variations [Fig. \ref{PT_TimeSeries}].

\subsection{Spectral analysis}
\begin{figure}[h]
\centering
\includegraphics[width=80mm]{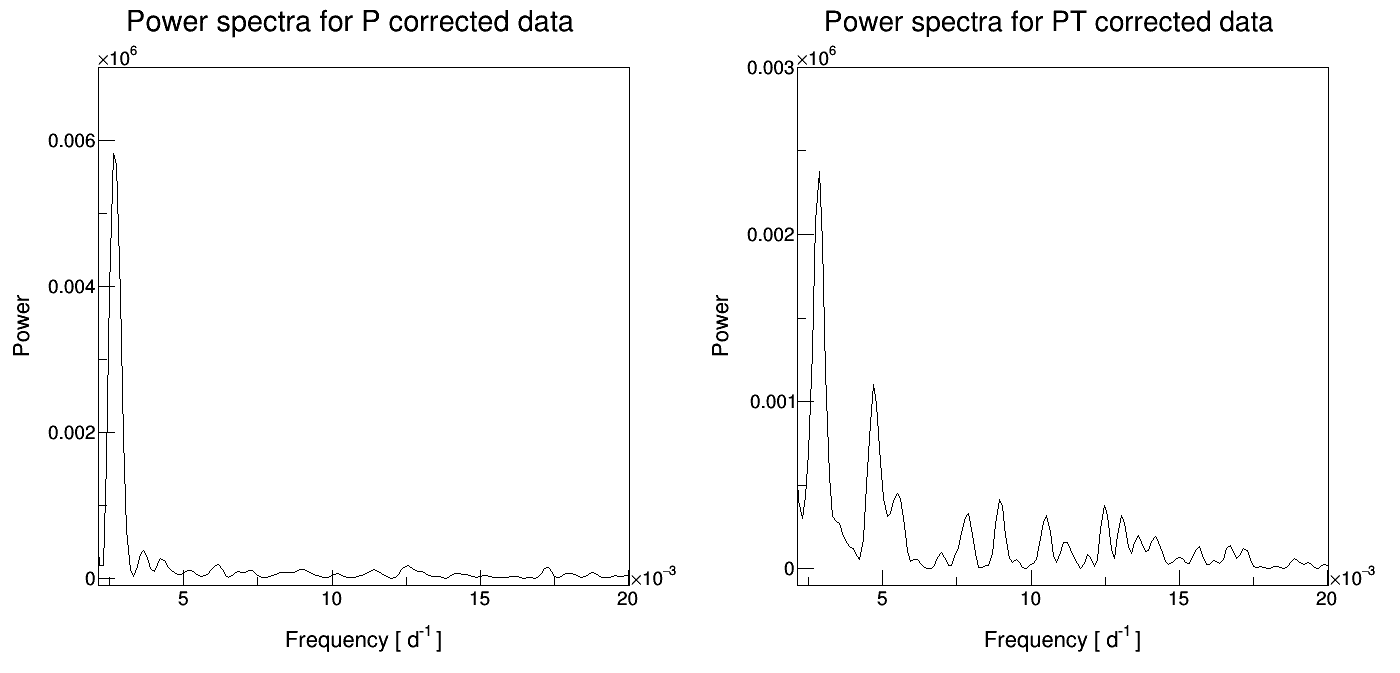}
\caption{Power spectra for pressure corrected and temperature and pressure corrected data.} 
\label{Spectral}
\end{figure}

Spectral analysis can give us more insight into effect of temperature correction on annual variation of muon count (presented for GLL data in [Fig. \ref{Spectral}])

After temperature correction, peak related to annual periodicity in power spectrum appears to be significantly reduced relative to nearby peaks.

\subsection{Neutron monitor correlation}
\begin{figure}[h]
\centering
\includegraphics[width=80mm]{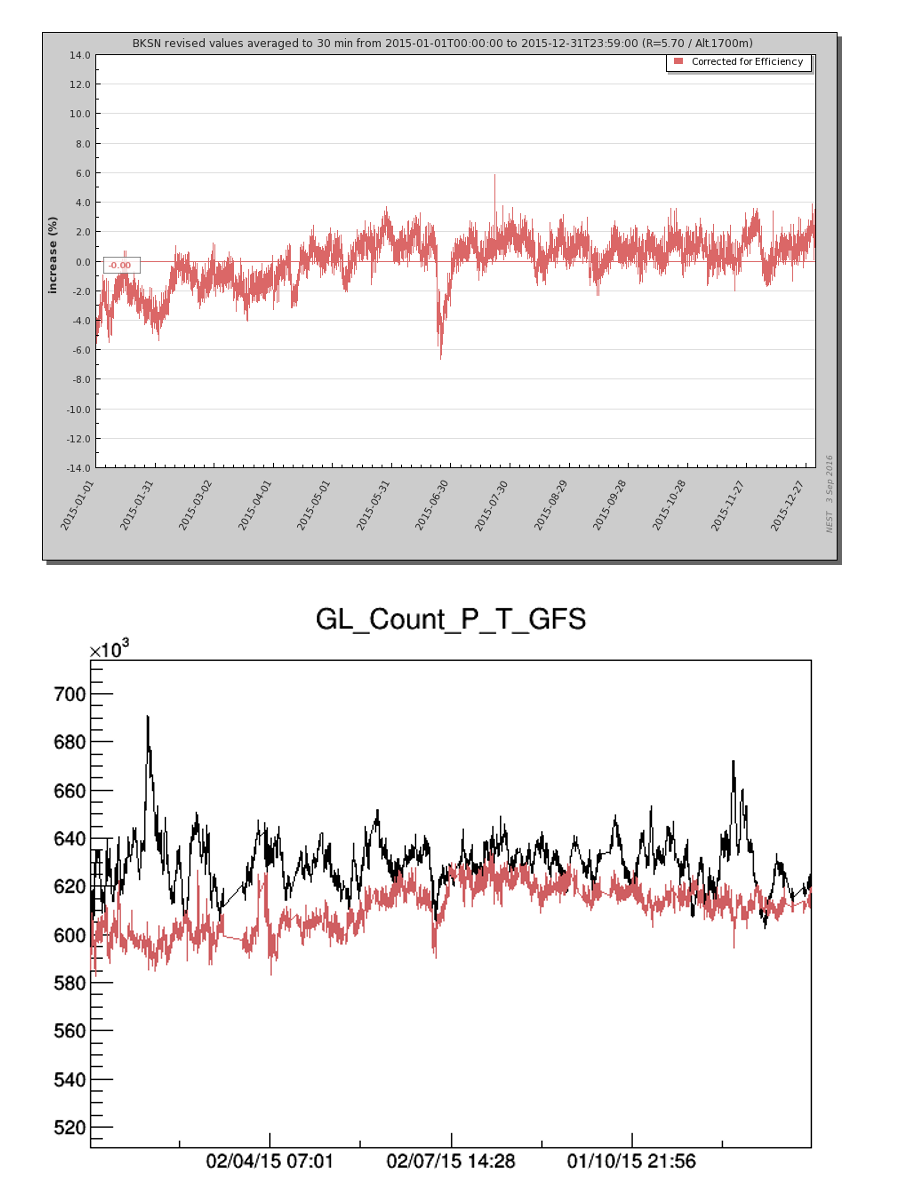}
\caption{BAKSAN neutron monitor (above) and GLL raw and pressure/temperature corrected data (below in red) comparison for year 2015.} 
\label{NM_Compare}
\end{figure}

Possible validation for correction procedure would be agreement of pressure/temperature 
corrected muon count time series with neutron monitor data. BAKSAN neutron monitor was 
selected as a possible reference [Fig. \ref{NM_Compare}].

\section{Conclusions}
Corrections for temperature and pressure effect are essential for muon data gathered at
Belgrade LBL. Atmospheric temperature profile for Belgrade seems to be adequately modeled by GFS. Temperature correction utilizing integral method seems to give acceptable results (while quality can still be further improved). Also, other methods could be applied and results compared. Muon flux data after pressure and temperature corrections has increased sensitivity to periodic and aperiodic effects of non-atmospheric origin. Preliminary comparison with neutron monitor data supports this claim with more detailed correlation analysis to follow in the future.

\section{Acknowledgements}
%\begin{acknowledgments}
The present work was funded by the Ministry of Education and Science of the Republic of Serbia, under the Project No. 171002. The Belgrade Laboratory bears  the  name  of  “Dr.  Radovan  Antanasijevic”, in honour of its early deceased founder and first director. 
%\end{acknowledgments}

\bigskip % extra skip inserted

\end{document}